\documentclass[pra,twocolumn,superscriptaddress]{revtex4}

\usepackage{psfrag}
\usepackage[T1]{fontenc}

\usepackage{amsfonts}
\usepackage{amssymb}

\usepackage{times}
\usepackage{pifont}
\usepackage{ae,aecompl}
\usepackage{amsmath}
\usepackage{graphicx}

\begin{document}

\title{Two-Dimensional Critical Potts and its Tricritical Shadow}
\author{Wolfhard Janke}
\affiliation{Institut f\"ur Theoretische Physik, Universit\"at Leipzig,
  Augustusplatz 10/11, 04109 Leipzig, Germany}
\author{Adriaan M. J. Schakel} \affiliation{Institut f\"ur Theoretische
Physik, Freie Universit\"at Berlin, Arnimallee 14, 14195 Berlin,
Germany}
\begin{abstract}
These notes give examples of how suitably defined geometrical objects
encode in their fractal structure thermal critical behavior.  The
emphasis is on the two-dimensional Potts model for which two types of
spin clusters can be defined.  Whereas the Fortuin-Kasteleyn clusters
describe the standard critical behavior, the geometrical clusters
describe the tricritical behavior that arises when including vacant
sites in the pure Potts model.  Other phase transitions that allow for a
geometrical description discussed in these notes include the superfluid
phase transition and Bose-Einstein condensation.
\end{abstract}

\date{\today}

\maketitle

\section{Introduction}

The quest for understanding phase transitions in terms of geometrical
objects has a long history.  One of the earlier examples, due to
Onsager, concerns the superfluid phase transition in liquid $^4$He---the
so-called $\lambda$ transition.  During the discussion of a paper
presented by Gorter, Onsager \cite{Onsager} made the following remark:
``As a possible interpretation of the $\lambda$-point, we can understand
that when the concentration of vortices reaches the point where they
form a connected tangle throughout the liquid, then the liquid becomes
normal.''  Feynman also worked on this approach and summarized the idea
as follows \cite{Feynman55}: ``The superfluid is pierced through and
through with vortex line.  We are describing the disorder of Helium~I.''
This approach focuses on vortex loops, i.e., one-dimensional geometrical
objects, which form a fluctuating vortex tangle.  As the critical
temperature $T_\lambda$ is approached from below, the vortex loops
proliferate and thereby disorder the superfluid state, causing the
system to revert to the normal state.  The $\lambda$ transition is thus
characterized by a fundamental change in the typical vortex loop size.
Whereas in the superfluid phase only a few small loops are present,
close to $T_\lambda$ loops of all sizes appear.  The sudden appearance
of arbitrarily large geometrical objects is reminiscent of what happens
in percolation phenomena at the percolation threshold where clusters
proliferate.  Even on an infinite lattice, a percolating cluster can be
found spanning the lattice.

\begin{figure}[hb!]
\begin{center}
\psfrag{1}[t][t][1][0]{$1$}
\psfrag{2}[t][t][1][0]{$2$}
\psfrag{3}[t][t][1][0]{$3$}
\psfrag{d}[t][t][1][0]{$d$}
\psfrag{x}[t][t][1][0]{$x$}
\psfrag{t}[t][t][1][0]{$\tau$}
\psfrag{h}[b][b][.8][0]{$\hbar/k_{\rm B} T$}
\includegraphics[width=.32\textwidth]{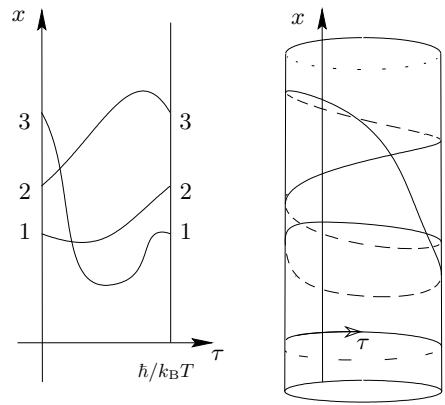}
\end{center}
\caption{The worldlines of three particles that, after moving a time
$\tau= \hbar/k_{\rm B} T$ in the imaginary time direction, are cyclically
permuted (left panel).  The three separate worldlines can also be
represented by a single worldline that winds three times around the
imaginary time axis (right panel). (After Ref.\ \protect\cite{loops}.)
\label{fig:path_labels}}
\end{figure}
A second example, due to Feynman \cite{Feynman53}, is related to
Bose-Einstein condensation. Here, the relevant geometrical objects are
worldlines.  In the imaginary-time formalism, used to describe
quantum systems at finite temperature $T$, the time dimension becomes
compactified, $t = - \mathrm{i} \tau$, with $0 \leq \tau \leq
\hbar/k_{\rm B} T$, where $k_{\rm B}$ is Boltzmann's constant.  Because
of periodic boundary conditions, the worldlines then form closed loops.
At high temperatures, where the system behaves more or less classically,
the individual particles form separate closed loops wrapping only once
around the imaginary time axis.  Upon lowering the temperature, these
small loops, describing single particles, hook up to form larger exchange
rings.  A particle in such a composite ring (see
Fig.~\ref{fig:path_labels}) moves in imaginary time along a trajectory
that does not end at its own starting position, but ends at that of
another particle.  Hence, although the initial and final configurations
are identical, the particles in a composite ring are cyclically permuted
and thus become indistinguishable \cite{Feynman53}.
Figure~\ref{fig:path_labels} gives an example of three particles,
labeled 1,2, and 3.  After wrapping once around the imaginary time axis
particle 1 ends at the starting position of particle 2, which in turn
ends after one turn around the imaginary time axis at the starting
position of particle 3.  That particle, finally, ends at the starting
position of particle 1.  In this way, the three particles are cyclically
permuted, forming the cycle $(1,2,3)$.  Being part of a single loop
which winds three times around the imaginary time axis, the particles
cannot be distinguished any longer.  At the critical temperature,
worldlines proliferate and---again as in percolation phenomena---loops
wrapping arbitrary many times around the imaginary time axis appear,
signaling the onset of Bose-Einstein condensation \cite{loops,percoPRE}.
This approach has been turned into a powerful Monte Carlo method by
Ceperley and Pollock \cite{CePo} that can even handle strongly
interacting systems like superfluid $^4$He (see Ref.\ \cite{Ceperley}
for a review).

\begin{figure}
\includegraphics[width=.25\textwidth]{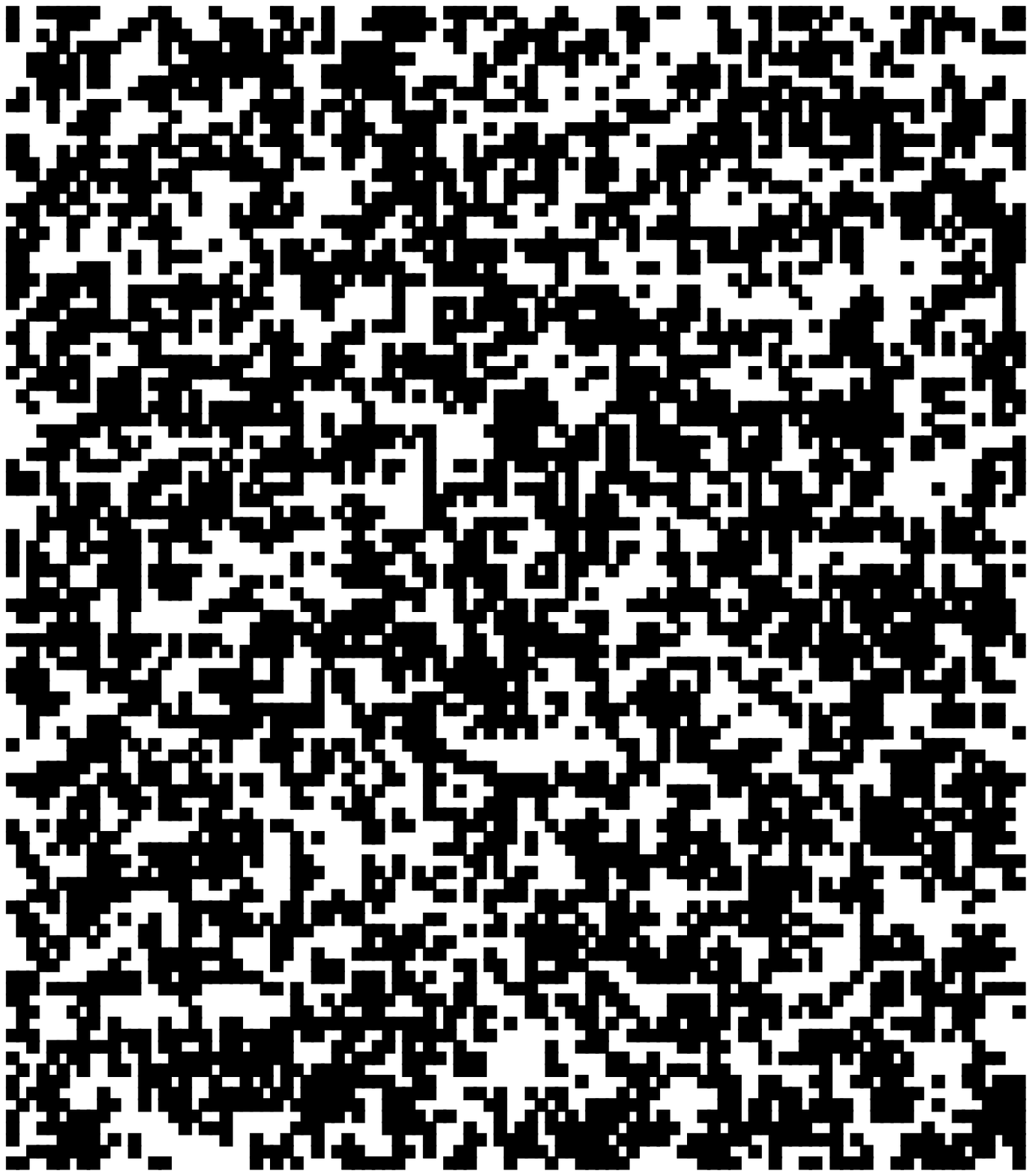} \hspace{-.8cm}
\includegraphics[width=.25\textwidth]{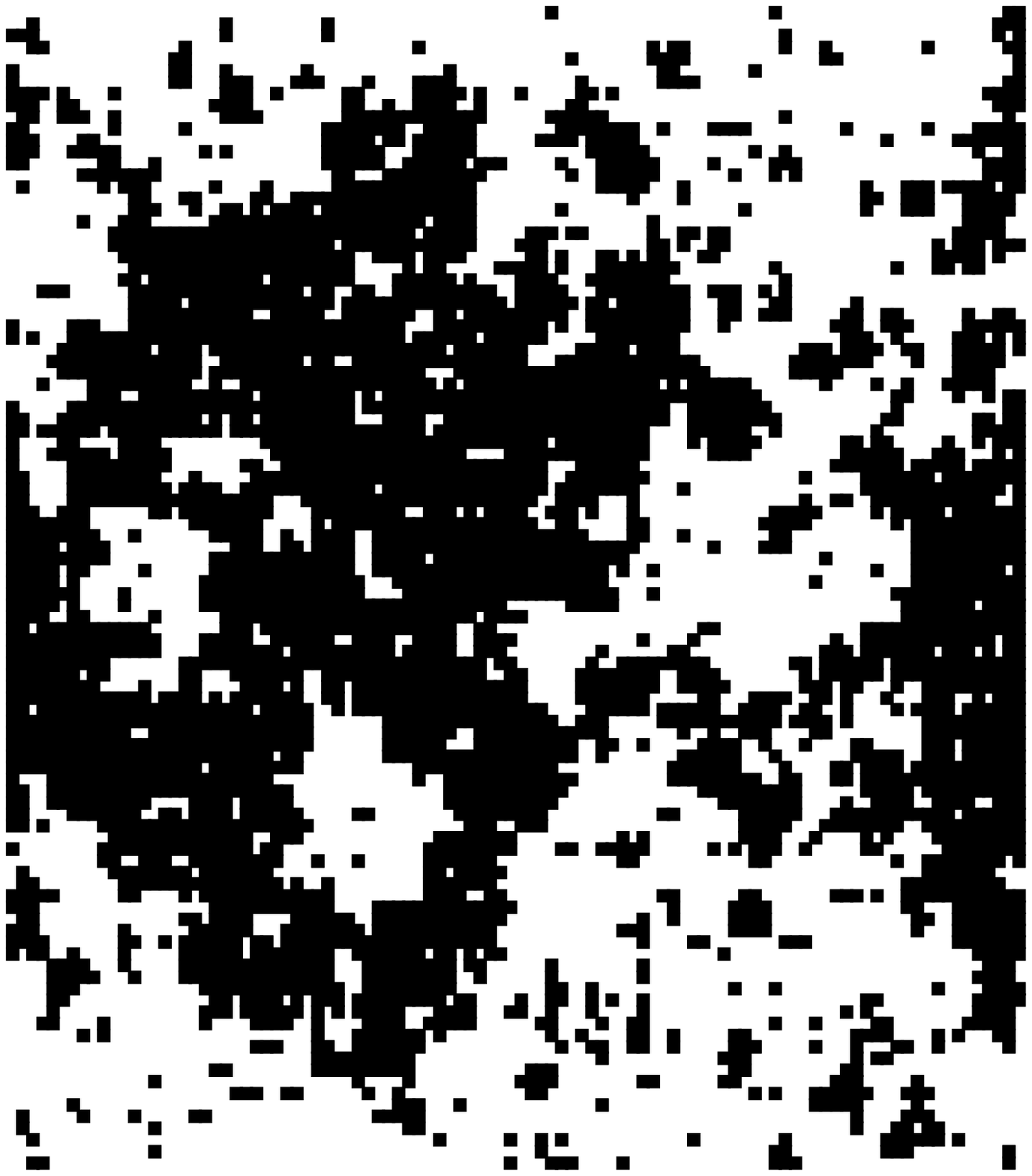}
\caption{Snapshots of typical spin configurations of the Ising model on
a square lattice of linear size $L=100$ in the normal, hot phase at
$\beta=0.5 \, \beta_\mathrm{c}$ (left panel) and just above the Curie
point at $\beta=0.98 \, \beta_\mathrm{c}$ (right panel). A spin up is
denoted by a black square, while a spin down is denoted by a white one.
\label{fig:ising}}
\end{figure}

A third example concerns the phase transition in simple magnets.  The
most elementary model describing such a transition is provided by the
Ising model, obtained by assigning a spin that can point either up or
down to each lattice site.  Figure \ref{fig:ising} shows typical spin
configurations for a square lattice in the normal, hot phase and just
above the Curie point. For convenience, a spin up is denoted by a black
square, while a spin down is denoted by a white one.  From these
snapshots, the relevant geometrical objects appear to be clusters of
nearest neighbor spins in the same spin state (in the following, we will
qualify this statement).  The normal, disordered phase consists of many
small clusters.  As the Curie point $T_\mathrm{c}$ is approached from
above, larger clusters appear, which at $T_\mathrm{c}$ start to
proliferate---as in percolation phenomena.  In the absence of an applied
magnetic field, the percolating cluster can consist of either up or down
spins, both having equal probability to form the majority spin state.
Since the percolating spin clusters have a fractal structure, it is
tempting to ask whether this structure encodes the standard
thermodynamic critical behavior, as in percolation theory?  More
generally, we wish to address in these notes the question: Can suitably
defined geometrical objects encode in their fractal structure the
standard critical behavior of the system under consideration?  To
highlight the basic features, we consider simple models, such as the
Ising, the Potts, and the XY model.  Moreover, we study them mostly in
two dimensions (2D) since many analytical predictions, obtained by using
Coulomb gas methods and conformal field theory, are available there.

The rest of these notes is organized as follows.  In the next section,
the 2D critical Potts model is discussed.  Central to the discussion is
the equivalent geometrical representation of this spin model in terms of
so-called Fortuin-Kasteleyn clusters \cite{FK}.  The fractal structure
of these stochastic clusters and the way the thermal critical behavior
of the Potts model can be extracted from it are studied in detail.  In
Sec.~\ref{sec:tri}, the tricritical Potts model is discussed.  The
clusters encoding the tricritical behavior turn out to be the naive
clusters of nearest neighbor spins in the same spin state, which feature
in Fig.~\ref{fig:ising}.  Their fractal structure is connected via a
dual map to that of the Fortuin-Kasteleyn clusters, which encode the
thermodynamic critical behavior.  In Sec.~\ref{sec:boundaries}, the
boundaries of both cluster types are studied.  The notes end with a
summary of the main results and an outlook to other applications.

\section{Critical Potts Model}
\subsection{Fortuin-Kasteleyn Representation}
The Potts model is one of the well studied spin models in statistical
physics \cite{Potts}.  It is defined by considering a lattice with each
lattice site given a spin variable $s_i=1,2,\cdots,Q$ that can take $Q$
different values.  In its standard form, the spins interact only with
their nearest neighbors specified by the Hamiltonian
\begin{equation} 
\mathcal{H} = - K\sum_{\langle ij\rangle} 
  \left(\delta_{s_i,s_j} - 1 \right),
\end{equation}  
where $K$ denotes the coupling constant.  Nearest neighbor spins notice
each other only when both are in the same spin state, as indicated by
the Kronecker delta.  The Potts model is of particular interest to us as
for $Q=2$ it is equivalent to the Ising model, while in the limit $Q \to
1$ it describes ordinary, uncorrelated percolation.  The notation
$\sum_{\langle ij\rangle}$ is to indicate that the double sum over the
lattice sites, labeled by $i$ and $j$, extends over nearest neighbors
only.  The partition function $Z$ can be written as
\begin{equation} 
\label{ZTr}
  Z = \mathrm{Tr} \; \mathrm{e}^{- \beta \mathcal{H}} = 
  \mathrm{Tr} \prod_{\langle ij\rangle} \left[(1-p) + p \, 
    \delta_{s_i,s_j} \right],
\end{equation} 
where $\beta$ denotes the inverse temperature, and the trace Tr stands
for the sum over all possible spin configurations.  In writing
Eq.~(\ref{ZTr}), use is made of the identity
\begin{equation}
\label{identity}
\mathrm{e}^{\beta \left(\delta_{s_i,s_j}-1\right)} = (1-p) + p \,
\delta_{s_i,s_j} ,
\end{equation} 
with $p = 1- \mathrm{e}^{-\beta}$, where here and in the sequel we 
set the coupling constant $K$ to unity.  The identity (\ref{identity})
can be pictured as setting bonds with probability $p/[(1-p) + p] = p$
between two nearest neighbor spins in the same spin state for which
$\delta_{s_i,s_j}=1$.  When two nearest neighbor spins are not in the
same spin state, $\delta_{s_i,s_j}=0$, then with probability
$(1-p)/(1-p) = 1$ the bond is not set, i.e., never.  It thus follows,
that the partition function can be equivalently written as
\begin{equation} 
\label{ZFK}
Z_{\rm FK} =  \sum_{\{\Gamma\}} p^b (1-p)^{\bar{b}+a} Q^{N_\mathrm{C}},
\end{equation} 
where $\{\Gamma\}$ denotes the set of bond configurations realized on a
total of $B$ bonds on the lattice.  A given configuration is specified
by $b$ set and $\bar{b}$ not set bonds between nearest neighbor spins in
the same spin state, and $a$ pairs of nearest neighbor spins not in the
same spin state (for which the bonds are never set).  Together they add
up to the total number of bonds, $B= b + \bar{b} + a$, so that the
exponent $\bar{b}+a$ in Eq.~(\ref{ZFK}) can also be written as $B-b$.
Only spins connected by set bonds form a cluster.  The exponent
$N_\mathrm{C}$ in Eq.~(\ref{ZFK}) denotes the number of clusters,
including isolated sites, contained in the bond configuration $\Gamma$.
The factor $Q^{N_\mathrm{C}}$ arises because a given cluster can be in
any of the $Q$ possible spin states.  Equation (\ref{ZFK}) is the
celebrated Fortuin-Kasteleyn (FK) representation of the Potts model
\cite{FK}.  It gives an equivalent representation of that spin model in
terms of FK clusters obtained from the naive geometrical clusters of
nearest neighbor spins in the same spin state, discussed in the
Introduction, by putting bonds with a probability $p = 1-
\mathrm{e}^{-\beta}$ between nearest neighbors.  As geometrical clusters
are split up in the process, the resulting FK clusters are generally
smaller and more loosely connected than the geometrical ones.

Not only does the FK representation provide a geometrical description of
the phase transition in the Potts model, it also forms the basis of
efficient Monte Carlo algorithms by Swendsen and Wang
\cite{SwendsenWang}, and by Wolff \cite{Wolff}, in which not individual
spins are updated, but entire FK clusters.  The main advantage of the
nonlocal cluster update over a local spin update, like Metropolis or
heat bath, is that it substantially reduces the critical slowing down
near the critical point. 

\subsection{FK Clusters}
The results of standard percolation theory \cite{StauferAharony} also
apply to FK clusters.  In particular, the distribution $\ell_n$ of FK
clusters, giving the average number density of clusters of mass $n$,
takes near the critical point the asymptotic form
\begin{equation} 
\label{elln}
\ell_n \sim n^{- \tau} \, \mathrm{e}^{- \theta \, n}. 
\end{equation} 
The first factor, characterized by the exponent $\tau$, is an entropy
factor, measuring the number of ways a cluster of mass $n$ can be
embedded in the lattice.  The second factor is a Boltzmann weight which
suppresses large clusters when the parameter $\theta$ is finite.
Clusters proliferate and percolate the lattice when $\theta$ tends to
zero.  The vanishing is characterized by a second exponent $\sigma$
defined via
\begin{equation} 
\label{theta}
\theta \propto |T-T_{\rm c}|^{1/\sigma}.
\end{equation}   
As in percolation theory \cite{StauferAharony}, the values of the two
exponents specifying the cluster distribution uniquely determine the
critical exponents.  To obtain these relations, we start by considering
the radius of gyration $R_n$,
\begin{equation} 
R_n^2 = \frac{1}{n} \sum_{i=1}^n ({\bf x}_i - \bar{\bf x})^2 =
\frac{1}{2n^2} \sum_{i,j=1}^n ({\bf x}_i - {\bf x}_j)^2,
\end{equation} 
with ${\bf x}_i$ the position vectors of the sites and $\bar{\bf x} =
(1/n) \sum_{i=1}^n {\bf x}_i$ the center of mass of the cluster. 
Asymptotically, the average $\langle R_n \rangle$ scales with the
cluster mass $n$ as
\begin{equation} 
\label{Hausdorff}
\langle R_n \rangle \sim n^{1/D},
\end{equation} 
which defines the Hausdorff, or fractal dimension $D$.  The average
radius of gyration $\langle R_n \rangle$ gives the typical linear size
of a cluster of mass $n$.  A second length scale is provided by the
correlation length $\xi$, which diverges close to $T_\mathrm{c}$ with an
exponent $\nu$ as $\xi \sim |T - T_\mathrm{c}|^{-\nu}$.  Both are
related via
\begin{equation} 
\langle R_n \rangle = \xi \, {\sf R}(n \theta),
\end{equation} 
where ${\sf R}$ is a scaling function, cf.\ Eq.~(\ref{elln}).  From the
asymptotic behavior (\ref{Hausdorff}), the divergence of the correlation
length, and the vanishing (\ref{theta}) of the parameter $\theta$ as
$T_\mathrm{c}$ is approached, the relation
\begin{equation}
\label{nu} 
\nu = \frac{1}{\sigma D}
\end{equation} 
follows, connecting the critical exponent $\nu$ to the fractal dimension
$D$ of the clusters and $\sigma$.  

The fractal dimension can also be related to the entropy exponent $\tau$
as follows.  At criticality, the mass $n$ of a cluster is distributed
over a volume of typical linear size $\langle R_n \rangle$, so that
\begin{equation} 
n \ell_n \sim 1/\langle R_n \rangle^d,
\end{equation} 
with $d$ the dimension of the lattice.  This leads to the well-known
expression
\begin{equation}
\label{tau} 
\tau = \frac{d}{D} + 1, 
\end{equation}  
in terms of which the correlation length exponent reads $\nu = (\tau
-1)/d\sigma$.
\subsection{Improved Estimators}
To see how physical observables, such as the magnetization $m$ and the
magnetic susceptibility $\chi$ are represented in terms of FK clusters,
we consider the Ising model in the standard notation with the spin
variable $S_i = \pm 1$ for simplicity.  

The correlation function $\langle S_i S_j \rangle$ has a particular
simple representation.  When the two spins belong to two different FK
clusters
\begin{equation} 
\langle S_i S_j \rangle = \frac{1}{4} \sum_{S_i, S_j=\pm1}  S_i S_j = 0,
\end{equation} 
while when they belong to the same cluster
\begin{equation} 
\langle S_i S_j \rangle = \frac{1}{2} \sum_{S_i=S_j=\pm1}  S_i S_j = 1.
\end{equation} 
That is, if $C_i$ denotes the FK cluster to which the spin $S_i$ belongs
and $C_j$ the one to which $S_j$ belongs, then
\begin{equation} 
\label{connec}
\langle S_i S_j \rangle = \delta_{C_i,C_j}.
\end{equation} 
For the susceptibility $\chi \equiv \sum_{i j} \langle S_i S_j
\rangle$ in the normal phase, Eq.~(\ref{connec}) gives
\begin{equation} 
\chi = \sum_{i j} \delta_{C_i,C_j} =  \sum_{\{C\}} n_C^2 ,
\end{equation} 
where the sum $\sum_{\{C\}}$ is over all FK clusters, and $n_C$ denotes
the mass of a given cluster.  In terms of the FK cluster distribution
$\ell_n$, the susceptibility can be written as
\begin{equation} 
\label{chiFK}
\chi = \sum_n n^2 \ell_n.
\end{equation} 
Note that in percolation theory \cite{StauferAharony}, the ratio $\sum_n
n^2 \ell_n/\sum_n n \, \ell_n$ denotes the average cluster size.  Since
in the Ising model all $L^d$ spins are part of some FK cluster, we have
the constraint
\begin{equation} 
\label{constraint}
\sum_n n \, \ell_n =1.
\end{equation} 
It thus follows that the right hand of Eq.~(\ref{chiFK}) precisely gives
the average size of FK clusters.  In other words, this geometrical
observable directly measures the magnetic susceptibility of the Ising
model.  From the asymptotic form (\ref{elln}), and the divergence $\chi
\sim |T-T_{\rm c}|^{-\gamma}$ of the susceptibility when the critical
point is approached, the relation $\gamma = (3-\tau)/\sigma$ between the
critical exponent $\gamma$ and the cluster exponents $\sigma$ and $\tau$
follows.

Also the magnetization $m$ has a simple geometrical representation
\cite{StauferAharony}.  In an applied magnetic field $H$, a spin cluster
of mass $n_C$ has a probability $\propto \exp(\beta n_C H)$ to be
oriented along the field direction, and a probability $\propto
\exp(-\beta n_C H)$ to be oriented against the field direction.  The
difference between these probabilities gives the magnetization $m_C$ per
spin in the cluster,
\begin{equation} 
m_C = \tanh (\beta n_C H).
\end{equation} 
Close to the critical temperature and in the thermodynamic limit $L^d
\to \infty$, the largest cluster dominates, and $\tanh (\beta
n_\mathrm{max} H) \to \pm 1$ for this cluster, depending on its
orientation.  The magnetization of the entire system (per spin) then
becomes
\begin{equation} 
\label{MP}
m = \pm P_\infty,
\end{equation} 
where $P_\infty=n_\mathrm{max}/L^d$ gives the fraction of spins in the
largest cluster---the so-called percolation strength.  Because of the
constraint (\ref{constraint}), it is related to the FK cluster
distribution via
\begin{equation}
\label{Pinfty} 
P_\infty = 1 - \left.\sum_n \right.' n \, \ell_n,
\end{equation} 
where the prime on the sum indicates that the largest FK spin cluster is
to be excluded.  The magnetization vanishes near the critical point as
$m \sim |T-T_{\rm c}|^\beta$.  Together with the asymptotic behavior of
the cluster distribution, Eq.~(\ref{MP}) with Eq.~(\ref{Pinfty}) gives
the relation $\beta = (\tau -2)/\sigma$.

These geometrical observables (average cluster size and percolation
strength) are called \textit{improved estimators} because they usually
have a smaller standard deviation than the spin observables.

The results just derived for the Ising model also apply to the rest of
the critical Potts models \cite{FK}.  In this way, the thermal critical
exponents of these models are completely determined by the exponents
$\sigma$ and $\tau$, characterizing the FK cluster distribution.
Specifically,
\begin{align} 
\label{perce}
\alpha & = 2 - \frac{\tau -1}{\sigma}, & \beta
&= \frac{\tau -2}{\sigma}, & \gamma & =
\frac{3-\tau}{\sigma}, \nonumber \\ \eta &= 2 +
d \frac{\tau-3}{\tau-1}, & \nu & = \frac{\tau
-1}{d \sigma},
\end{align}
as in percolation theory \cite{StauferAharony}.  The exponent $\eta$,
determining the algebraic decay of the correlation function at the
critical point, is related to the fractal dimension via
\begin{equation} 
\label{Deta}
D = \tfrac{1}{2} (d+2-\eta). 
\end{equation} 
Consequently
\begin{equation} 
\label{gammanu}
\gamma/\nu = 2 D -d. 
\end{equation} 

\subsection{Critical Exponents}
The critical exponents of the 2D $Q$-state Potts model are known exactly
\cite{denNijs}.  It is convenient to parametrize the models as
\begin{equation}
\label{Potts_branch} 
\sqrt{Q} = - 2 \cos(\pi/\bar\kappa), 
\end{equation} 
with $2 \ge \bar\kappa \ge 1$.  For the Ising model ($Q=2$)
$\bar\kappa=4/3$, while for uncorrelated percolation ($Q \to 1$)
$\bar\kappa=3/2$.  The correlation length exponent $\nu$ and the
exponent $\eta$ are given in this representation by \cite{denNijs}:
\begin{equation}
\label{cesPotts} 
\frac{1}{\nu} = y_{\mathrm{T},1} = 3 -\frac{3}{2} \bar\kappa, \quad \eta = 2 -
\frac{1}{\bar\kappa} - \frac{3}{4} \bar\kappa,
\end{equation}
where $y_{\mathrm{T},1}$ is the leading thermal exponent.  The
next-to-leading thermal exponent $y_{\mathrm{T},2}$ reads
$y_{\mathrm{T},2}=4(1-\bar \kappa)$, which is negative for $\bar\kappa
\ge 1$, implying that the corresponding operator is an irrelevant
perturbation.  The other critical exponents can be obtained through
standard scaling relations. The parameter $\bar\kappa$ is related to the
central charge $c$, defining the universality class, via \cite{Cardyrev}
\begin{equation} 
\label{conf}
c = 1 - \frac{6(1-\bar\kappa)^2}{\bar\kappa}. 
\end{equation} 
Finally, the fractal dimension $D$ of FK clusters is given by
\cite{Stanley,Coniglio1989} 
\begin{equation} 
D = 1 + \frac{1}{2\bar\kappa} + \frac{3}{8} \bar\kappa ,
\end{equation} 
which gives $D=15/8$ for the Ising model and $D=91/48$ for uncorrelated
percolation. 

\begin{figure}
\begin{center}
\includegraphics[width=.4\textwidth]{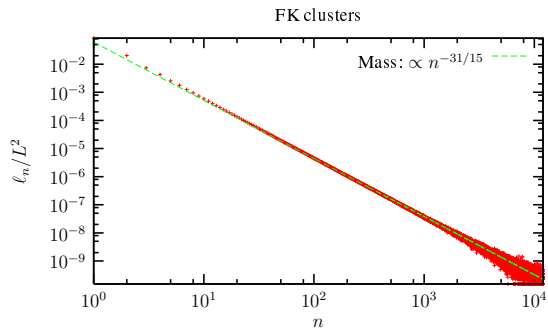} 
\end{center}
\caption{Distribution $\ell_n$ normalized to the volume $L^2$ of FK
clusters in the 2D Ising model at criticality on a square lattice of
linear size $L=512$.  Statistical error bars are omitted from the data
points for clarity.  The straight line is a one-parameter fit through
the data points with (minus) the slope fixed to the predicted value
$\tau=31/15=2.06667$.  The fit illustrates that asymptotically the
distribution is algebraic, as expected at criticality. 
  \label{fig:dis_SW_clusternc_512}}
\end{figure}

To demonstrate that FK clusters actually percolate at the critical
point, Fig.~\ref{fig:dis_SW_clusternc_512} shows the distribution
$\ell_n$ of these clusters in the 2D Ising model at criticality
($\theta=0$) on a square lattice of linear size $L=512$.  With $D=15/8$,
it follows from Eq.~(\ref{tau}) that the entropy exponent takes the
value $\tau=31/15$.  The straight line, obtained through a one-parameter
fit with the slope fixed to the predicted value, shows that
asymptotically the FK cluster distribution has the expected behavior.

\subsection{Swendsen-Wang Cluster Update}
\begin{figure}
\begin{center}
\includegraphics[width=.4\textwidth]{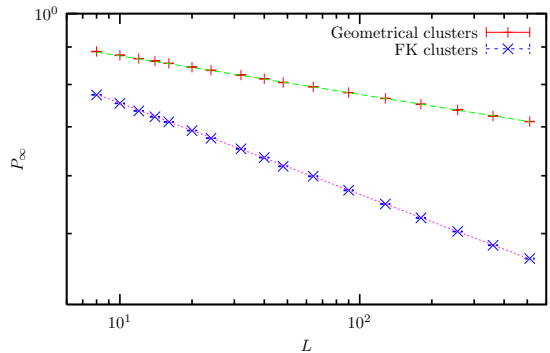} 
\end{center}
\caption{Log-log plot of the percolation strength $P_\infty$ of
geometrical and FK clusters at criticality in the 2D Ising model as a
function of the linear system size $L$.  The straight lines $0.988281 \,
L^{-0.0527}$ for geometrical and $1.00558 \, L^{-0.1248}$ for FK
clusters are obtained from two-parameter fits through the data points.
Statistical error bars are smaller than the symbol sizes.
  \label{fig:compare_beta}}
\end{figure}

The theoretical predictions (\ref{perce}) can be directly verified
through Monte Carlo simulations, using the Swendsen-Wang cluster update
\cite{SwendsenWang}.  Instead of single spins, entire FK clusters are
considered units to be flipped as a whole in this approach.  Standard
finite-size scaling theory applied to the percolation strength
$P_\infty$ and the average cluster size $\chi$ gives the scaling laws
\begin{equation} 
\label{finitess}
P_\infty = L^{-\beta/\nu} \, {\sf P}(L/\xi), \quad
\chi = L^{\gamma/\nu} \, {\sf X} (L/\xi),
\end{equation} 
with ${\sf P}$ and ${\sf X}$ scaling functions.  Precisely at $T_{\rm
c}$, these scaling relations imply an algebraic dependence on the system
size $L$, allowing for a determination of the exponent ratios $\beta/\nu$
(see Fig.~\ref{fig:compare_beta}) and $\gamma/\nu$.  Using these
geometrical observables as improved estimators for the magnetization and
susceptibility, respectively, we arrived at the estimates for the Ising
model ($Q=2$) \cite{fracIsing}
\begin{eqnarray} 
\beta/\nu &=& 0.1248(8) \approx 1/8, \nonumber \\ \gamma/\nu &=&
1.7505(12) \approx 7/4,
\end{eqnarray} 
where the right hands give the known values for the Ising critical
exponents.  These estimates illustrate first of all that FK clusters
indeed encode the thermal critical behavior of the Ising model.
Moreover, they also illustrate that measuring geometrical observables
gives excellent results for the critical exponents.  The data were
fitted over the range $L = 64 - 512$, using the least-squares
Marquardt-Levenberg algorithm.

In Ref.~\cite{AABRH}, the fractal dimension of FK clusters were obtained
from analyzing their distribution.  This method gives less accurate
results than applying finite-size scaling to improved estimators.  The
main problem is related to the fitting window.  The fitting range cannot
be started at too small cluster sizes, where the distribution has not
taken on its asymptotic form yet, while too large cluster sizes, which
are generated only a few times during a complete Monte Carlo run, are
also to be excluded because of the noise in the data and finite-size
effects.  The results depend sensitively on the precise choice of the
fitting window.

\subsection{Geometrical Clusters}
\begin{figure}
\begin{center}
\includegraphics[width=.4\textwidth]{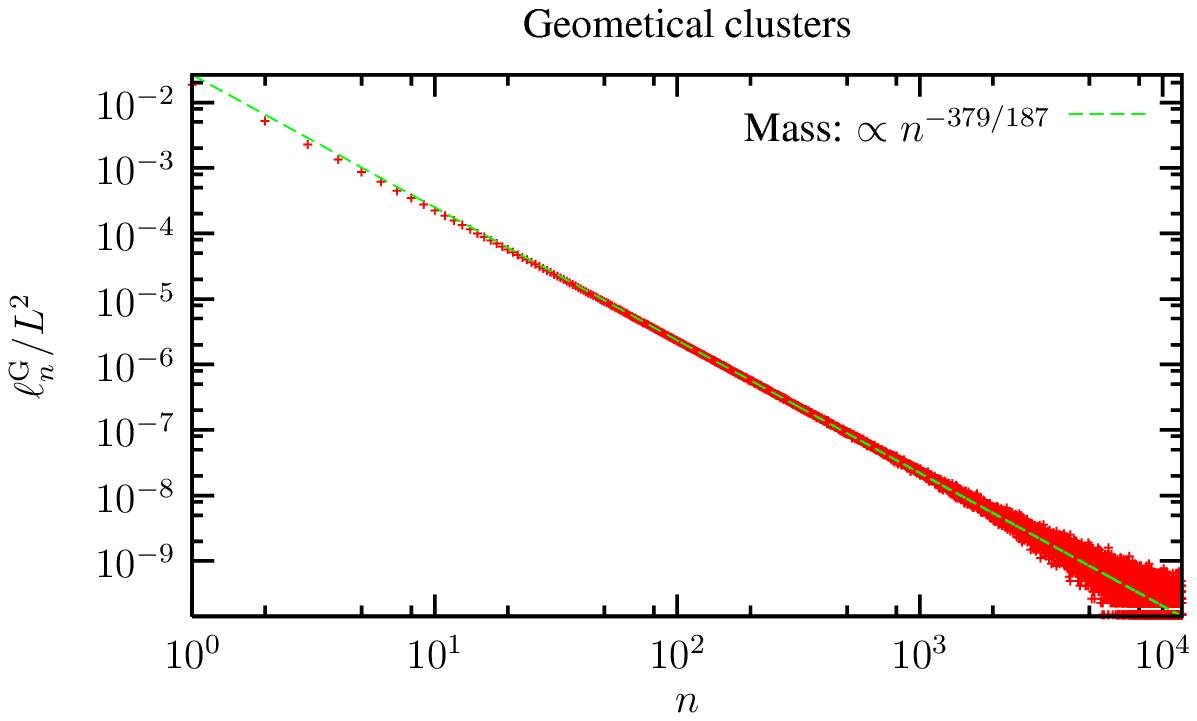} 
\end{center}
\caption{Distribution $\ell^\mathrm{G}_n$ normalized to the volume $L^2$
of geometrical clusters in the 2D Ising model at criticality on a square
lattice of linear size $L=512$.  Statistical error bars are omitted from
the data points for clarity.  The straight line is a one-parameter fit
through the data points with (minus) the slope fixed to the value
$\tau^\mathrm{G}=379/187= 2.02674$.  The fit illustrates that asymptotically
the distribution is algebraic at criticality. 
  \label{fig:dis_geo_clusternc_512}}
\end{figure}

Figure \ref{fig:ising} suggests that the geometrical spin clusters
also percolate right at the Curie point of the Ising model.  To
demonstrate this to be the case, Fig.~\ref{fig:dis_geo_clusternc_512}
shows the distribution of these clusters at criticality. 
Asymptotically, the distribution indeed shows algebraic behavior,
implying that clusters of all size appear in the system. 
It is therefore natural to investigate the exponents associated with the
percolation strength $P^\mathrm{G}_\infty$ (see
Fig.~\ref{fig:compare_beta}) and the average size $\chi^\mathrm{G}$ of
these geometrical clusters.  (The superscript ``G'' refers to
geometrical clusters.)  Using finite-size scaling, as for the FK
clusters, we arrived at the estimates \cite{fracIsing}
\begin{eqnarray} 
\label{betageo}
\beta^\mathrm{G}/\nu &=& 0.0527(4) \approx 5/96 = 0.0521 \cdots , \nonumber \\
\gamma^\mathrm{G}/\nu &=& 1.8951(5) \approx 91/48 = 1.8958 \cdots.
\end{eqnarray} 
In obtaining these estimates we included percolating clusters.  When
excluding them, as was done in Ref.~\cite{Fortunato}, the estimates
become less accurate \cite{fracIsing}.  
The entropy exponent which
follows from these results is $\tau^\mathrm{G}=379/187$, corresponding
to the fractal dimension $D^\mathrm{G}=187/96$.

It should be stressed that only in 2D geometrical clusters percolate
right at the critical temperature.  In higher dimensions, geometrical
clusters percolate in general too early at a lower temperature, and
their fractal structure is unrelated to any thermodynamic singularity.

The 2D exponents (\ref{betageo}) are not related to the critical
behavior of the Ising model and the question arises: What do these
exponents describe?

\section{Tricritical Potts Model}
\label{sec:tri}
\subsection{Dual Map}
\label{sec:dual}
When the pure 2D Potts model is extended to include vacant sites, it
displays in addition to critical also tricritical behavior at the same
critical temperature $T_\mathrm{c}$ \cite{NBRS}.  The tricritical
behavior is known to be intimately connected to the critical behavior,
and both critical points share the same central charge.  To demonstrate
this connection, note that for a given $c$, Eq.~(\ref{conf}) yields two
solutions for $\bar\kappa$:
\begin{equation} 
\label{kappac}
\bar\kappa_\pm = \frac{13-c \pm \sqrt{(c-25)(c-1)}}{12},
\end{equation}
with $\bar\kappa_+ \bar\kappa_- = 1$, where $\bar\kappa \equiv
\bar\kappa_+\ge1$ and hence $\bar\kappa_-\le1$.  Stated alternatively,
the substitution $\bar\kappa$ with $1/\bar\kappa$ leaves the central
charge (\ref{conf}) unchanged, $c({\bar\kappa}) = c(1/{\bar\kappa})$.
When applied to the parametrization (\ref{Potts_branch}) of the critical
Potts branch, this so-called \textit{dual map} yields the parametrization
\cite{Cardyrev}
\begin{equation} 
\label{partri}
\sqrt{Q^\mathrm{t}} = -2 \cos (\pi \bar\kappa),
\end{equation} 
of the tricritical branch (the superscript ``t'' refers to the
tricritical point).  Various results for the critical point
\cite{denNijs} can be simply transcribed to the tricritical point by
using this dual map, leading to \cite{NBRS,Nienhuis}
\begin{equation} 
\label{triexp}
\frac{1}{\nu^\mathrm{t}} = y^\mathrm{t}_{\mathrm{T},1} = 3 -\frac{3}{2
\bar\kappa}, \quad \eta^\mathrm{t} = 2 - \bar\kappa - \frac{3}{4
\bar\kappa},
\end{equation}
while the next-to-leading thermal exponent becomes
\begin{equation} 
\label{trisub}
y^\mathrm{t}_{\mathrm{T},2}=4- \frac{4}{\bar \kappa} . 
\end{equation} 
To preserve relation (\ref{Deta}) under the dual map, the fractal
dimensions of the geometrical and FK clusters must also be related by
the map $\bar \kappa \to 1/\bar \kappa$ \cite{geoPotts,DBN}. This gives 
\begin{equation} 
\label{Dgeo}
D^{\rm G} = 1 + \frac{\bar\kappa}{2} + \frac{3}{8\bar\kappa},
\end{equation} 
which is indeed the correct fractal dimension of geometrical clusters
\cite{StellaVdzandePRL,DS89}.  In other words, the geometrical clusters
can, as far as their scaling behavior is concerned, be considered
shadows of the FK clusters.  The use of the word ``shadow'' will become
clear when we consider the cluster boundaries in the next section.

\subsection{Ising \& its $Q^\mathrm{t} = 1$ Potts Shadow}
Equation~(\ref{Dgeo}) gives as fractal dimension of the geometrical
clusters of the Ising model ($\bar\kappa=4/3$) $D^\mathrm{G}=187/96$,
implying via Eq.~(\ref{tau}) $\tau= 379/187$, in accordance with what we
found numerically \cite{fracIsing}.  Note that with $\bar\kappa=4/3$,
Eq.~(\ref{partri}) gives $Q^\mathrm{t} = 1$.  That is, the tricritical
model described by the geometrical clusters of the Ising model is the
diluted $Q^\mathrm{t} = 1$ Potts model.  Both models share the same
central charge $c=1/2$.

The alert reader may have noticed a curiosity concerning the thermal
exponents.  According to Eq.~(\ref{triexp}), the correlation length
exponent $\nu^\mathrm{t}$ takes the value $\nu^\mathrm{t} = 1/
y^\mathrm{t}_{\mathrm{T},1}= 8/15$ in the diluted $Q^\mathrm{t} = 1$
Potts model ($\bar\kappa=4/3$).  Yet, in our numerical investigation
\cite{fracIsing} of the geometrical clusters of the Ising model, we seem
to observe the correlation length exponent $\nu=1$ of the Ising model.
Hence, $\nu$ and not the tricritical exponent $\nu^\mathrm{t}$ appears
in Eq.~(\ref{betageo}).  In fact, what we see is the tricritical
next-to-leading thermal exponent (\ref{trisub}), which for the diluted
$Q^\mathrm{t} = 1$ Potts model happens to take the same value as the
leading thermal exponent of the Ising model,
$y^\mathrm{t}_{\mathrm{T},2} = y_{\mathrm{T},1} = 1$ for
$\bar\kappa=4/3$.

\section{Hulls \& External Perimeters}
\label{sec:boundaries}
\subsection{FK Clusters}
When clusters percolate at a certain threshold, their boundaries
necessarily do too.  In the context of uncorrelated percolation in 2D,
external cluster boundaries can be traced out by a biased random walker
as follows \cite{GrossmanAharony}.  The algorithm starts by identifying
two endpoints on a given cluster, and putting the random walker at the
lower endpoint.  The walker is instructed to first attempt to move to
its nearest neighbor to the left.  If that site is vacant, the walker
should try to move straight ahead.  If that site is also vacant, the
walker should try to move to its right.  Finally, if also that site is
vacant, the walker is instructed to return to the previous site, to
discard the direction already explored, and to investigate the (at most
two) remaining directions in the same order.  When turning left or
right, the walker changes its orientation accordingly.  The procedure is
repeated iteratively until the upper endpoint is reached.  The other
half of the boundary is obtained by repeating the entire algorithm for a
random walker instructed to first attempt to move to its right rather
than to its left.

\begin{figure}
\begin{center}
\psfrag{a}[t][t][1][0]{(a)}
\psfrag{b}[t][t][1][0]{(b)}
\includegraphics[width=.4\textwidth]{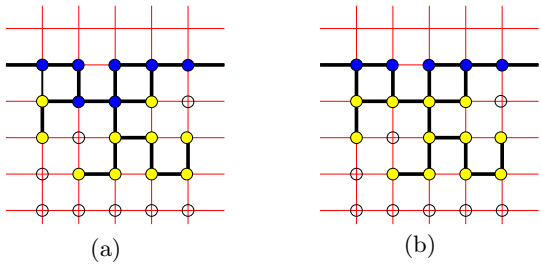}
\end{center}
\caption{In both panels, a piece of the same single FK cluster of
  nearest neighbor sites (filled circles) connected by bonds (black
  links) is shown.  Two different external boundaries can be defined:
  (a) The \textit{hull} (dark filled circles) is found by allowing a
  random walker tracing out the boundary to move only over set bonds.
  (b) The \textit{external perimeter} (dark filled circles) is found by
  allowing the random walker to move to a nearest neighbor on the
  cluster boundary irrespective of whether the connecting bond is set or
  not.  The external perimeter, which contains two sites less than the
  hull for this boundary segment, is therefore a smoother version of the
  hull.
  \label{fig:rw_algo_2}}
\end{figure}

For FK clusters, being built from bonds between nearest neighbor sites
with their spin in the same spin state, one can imagine two different
external boundaries (see Fig.~\ref{fig:rw_algo_2}).  First, one can
allow the random walker to move along the FK boundary only via set
bonds.  This defines the \textit{hull} of the cluster.  Second, one can
allow the random walker to move to a nearest neighbor site on the FK
boundary irrespective of whether the bond is set or not.  This defines
the \textit{external perimeter} of the cluster, which is a smoother
version of the hull.
Since boundaries are clusters themselves, they too are characterized by
a fractal dimension and a distribution like Eq.~(\ref{elln})
specified by two exponents $\sigma$ and $\tau$.  

\subsection{Fractal Dimensions}
The fractal dimensions of the hulls (H) and external perimeters (EP) of
FK clusters are given by \cite{SD,Coniglio1989,Duplantier00} 
\begin{equation} 
\label{dhdep}
D_\mathrm{H} = 1 + \frac{\bar\kappa}{2}, \quad D_\mathrm{EP} = 1 +
\frac{1}{2\bar\kappa}. 
\end{equation} 
As for clusters, the average hull and external perimeter sizes diverge
at the percolation threshold.  Let $\gamma_\mathrm{H}$ and
$\gamma_\mathrm{EP}$ denote the corresponding exponents, then because of
Eq.~(\ref{gammanu}) with $d=2$ and Eq.~(\ref{dhdep})
\begin{equation} 
\gamma_\mathrm{H}/\nu = \bar\kappa, \quad
\gamma_\mathrm{EP}/\nu = 1/\bar\kappa,
\end{equation} 
where a single correlation length exponent $\nu$ is assumed. 

\begin{figure}
\begin{center}
\includegraphics[width=.4\textwidth]{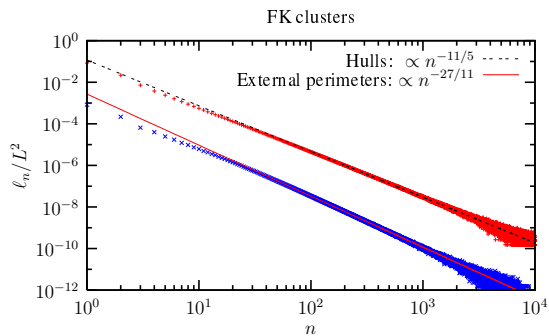}
\end{center}
\caption{Distribution normalized to the volume $L^2$ of the hulls and
external perimeters of FK clusters in the 2D Ising model at criticality
on a square lattice of linear size $L=512$.  Statistical error bars are
omitted from the data points for clarity.  The straight lines are
one-parameter fits through the data points with the slopes fixed to the
expected values.  For clarity, the external perimeters distribution is
shifted downward by two decades. 
  \label{fig:dis_SWnc_512_2}}
\end{figure}

For illustrating purposes, Fig.~\ref{fig:dis_SWnc_512_2} shows the
distribution of the two boundaries of FK clusters in the Ising model at
criticality.  The straight lines are one-parameter fits through the data
points with the slopes fixed to the expected values.  Although the
estimates for $D_\mathrm{H}$ and $D_\mathrm{EP}$ we obtained, using
finite-size scaling applied to the improved estimators at criticality,
are compatible with the theoretical conjectures \cite{geoPotts}, the
achieved precision is less than the one we reached for the clusters
themselves.  The reason for this is as follows.  While including
percolating clusters when considering the mass of the clusters, we
ignore them in tracing out cluster \textit{boundaries}.  Because of the
finite lattice size, large percolating clusters have anomalous small
(external) boundaries, so that including them would distort the boundary
distributions.  Moreover, the Grossman-Aharony algorithm
\cite{GrossmanAharony} used to trace out cluster boundaries generally
fails on a percolating cluster as its boundary not necessarily forms a
single closed loop any longer.  However, as we explicitly demonstrated
for the cluster mass \cite{fracIsing}, disregarding percolating clusters
leads to strong corrections to scaling, and therefore to less accurate
results.

\subsection{Geometrical Clusters}
For geometrical clusters, where the bond between nearest neighbor sites
with their spin in the same spin state is so to speak always set, hulls
and external perimeters cannot be distinguished, and
\begin{equation} 
D^\mathrm{G}_\mathrm{H} =  D^\mathrm{G}_\mathrm{EP}. 
\end{equation} 
The fractal dimension of the boundary is gotten from that of the hull
(\ref{dhdep}) of FK clusters by applying the dual map $\bar \kappa \to
1/\bar \kappa$, yielding \cite{Vanderzande} 
\begin{equation} 
D^\mathrm{G}_\mathrm{H} = 1 + \frac{1}{2\bar\kappa}. 
\end{equation} 
Since FK clusters have two boundaries, while geometrical clusters have
only one, geometrical clusters have less structure and can be considered
shadows of FK clusters under the dual map, as far as their scaling
behavior is concerned. 

\begin{figure}
\begin{center}
\includegraphics[width=.4\textwidth]{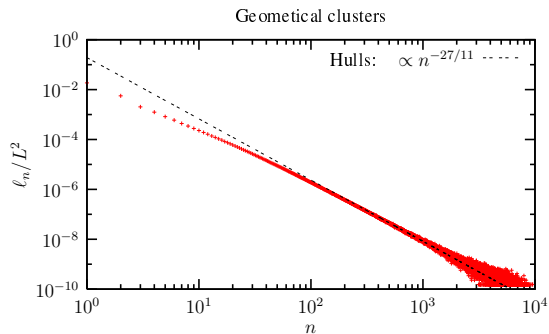}
\end{center}
\caption{Distribution normalized to the volume $L^2$ of the hulls
geometrical clusters in the 2D Ising model at criticality on a square
lattice of linear size $L=512$.  Statistical error bars are omitted from
the data points for clarity.  The straight line is a one-parameter fit
through the data points with the slope fixed to the expected value. 
  \label{fig:dis_geonc_512_2}}
\end{figure}

Again for illustrating purposes, Fig.~\ref{fig:dis_geonc_512_2} shows
the distribution of the hulls of geometrical clusters in the Ising model
at criticality.  The slow approach to the asymptotic form, with the
associated strong corrections to scaling we observed for the hulls of
geometrical clusters, stands out clearly from the other distributions.
The reason for this is that geometrical clusters have a larger extent
than FK clusters.  On a finite lattice, percolating clusters gulp up
smaller ones reached by crossing lattice boundaries.  For geometrical
clusters this happens more often than for FK clusters, so that
disregarding percolating clusters when tracing out cluster boundaries
has a more profound effect.  In particular, the average hull size is
underestimated.  With increasing lattice size, the effect becomes
smaller, as we checked explicitly \cite{fracIsing}.

This behavior is different from what we found in another numerical study
of the hulls of geometrical clusters in the 2D Ising model
\cite{geoPotts}.  In that Monte Carlo study, we used a plaquette update
to directly simulate the hulls.  Although the largest hull was omitted
in each measurement, we found corrections to scaling to be virtually
absent (see Fig.~11 of that paper).  This allowed us to obtain a precise
estimate for the fractal dimension on relatively small lattices.

\section{Conclusions \& Outlook}
As illustrated in these notes, for the 2D Potts models it is well
established that suitably defined geometrical objects encode in their
fractal structure critical behavior.  In fact, two types of spin
clusters exist, viz., FK and geometrical clusters, which both
proliferate precisely at the thermal critical point.  As emphasized
before, this is special to 2D.  In general, geometrical clusters
percolate at an inverse temperature $\beta_\mathrm{p} >
\beta_\mathrm{c}$.  The fractal structure of FK clusters encodes the
critical exponents of the critical Potts model, while that of
geometrical clusters in 2D encodes those of the tricritical Potts model.
The fractal structure of the two cluster types as well as the two fixed
points are closely related, being connected by the dual map $\bar \kappa
\to 1/\bar \kappa$.  This map conserves the central charge, so that both
fixed points share the same central charge.  The geometrical clusters
can, as far as scaling properties are concerned, be considered shadows
of the FK clusters.

Up to now we considered external boundaries of spin clusters as clusters
themselves, which necessarily percolate when the spin clusters do.  An
alternative way of looking at these boundaries is to consider them as
loops.  In this approach, it is natural to extend the Ising model in
another way and to consider the O($N$) spin models, with $-2\leq N \leq
2$.  The high-temperature (HT) representation of the critical O($N$)
spin model naturally defines a loop gas, corresponding to a diagrammatic
expansion of the partition function in terms of closed graphs along the
bonds on the underlying lattice \cite{Stanley_book}.  The loops
percolate right at the critical temperature, and similar arguments as
given in these notes for spin clusters show that the fractal structure
of these geometrical objects encode important information concerning the
thermal critical O($N$) behavior \cite{ht,comment}.  This connection was
first established by de Gennes \cite{deGennes} for self-avoiding walks,
which are described by the O($N$) model in the limit $N \to 0$.  One
aspect in which lines differ from spin clusters is that they can be open
or closed.  It is well known from the work on self-avoiding walks that
the loop distribution itself is not sufficient to establish the critical
behavior, as has recently also been emphasized in
Ref.~\cite{ProkofevSvistunov}.  For this, also the total number $z_n
\equiv \sum_j z_n(\mathbf{x}_i, \mathbf{x}_j)$ of \textit{open graphs}
of $n$ steps starting at $\mathbf{x}_i$ and ending at an arbitrary site
$\mathbf{x}_j$ is needed.  Its asymptotic behavior close to the critical
temperature, cf.\ Eq.~(\ref{elln}) with Eq.~(\ref{tau}),
\begin{equation}
\label{zn}
  z_n \sim n^{\vartheta/D} {\rm e}^{- \theta n},
\end{equation}   
provides an additional exponent $\vartheta$, which together with the
loop distribution exponents is needed to specify the full set of
critical exponents \cite{comment}.  In Eq.~(\ref{zn}), $D$ denotes the
fractal dimension of the closed graphs.  Note that for spin clusters, the
notion of open or closed does not apply, so that the analog of the
exponent $\vartheta$ is absent there.

Remarkably, the HT graphs of a given critical O($N$) model represent at
the same time the hulls of the geometrical clusters in the $Q$-state
Potts model with the same central charge
\cite{VdzandeStellaJP,DS89,geoPotts}. To close the circle, we note that,
as in the Potts model, including vacancies in the O($N$) model gives
rise to also tricritical behavior.  The tricritical point corresponds to
the point where the HT graphs collapse.  In the context of self-avoiding
walks ($N \to 0$), this point is known as the $\Theta$ point.  Using the
duality discussed in Sec.~\ref{sec:dual}, we recently conjectured that
the tricritical HT graphs at the same time represent the hulls of the FK
clusters of the Potts model with the same central charge $c$ as the
tricritical O($N$) model \cite{ht}.  This connection allowed us to
predict the magnetic scaling dimension of the O($N)$ tricritical model,
in excellent agreement with recent high-precision Monte Carlo data in
the range $0 \leq c \lesssim 0.7$ \cite{bloeteetal}.

We started these notes mentioning the $\lambda$ transition in liquid
$^4$He in terms of vortex proliferation.  In closing, we wish to give
the present status of that picture as established in a very recent
high-precision Monte Carlo study of the 3D complex $|\phi|^4$ theory
describing the transition \cite{vortexline}. An important observable is
the \textit{total} vortex line density $v$.  By means of standard
finite-size scaling analysis of the associated susceptibility $\chi =
L^3(\langle v^2 \rangle - \langle v \rangle^2)$, the inverse critical
temperature $\beta_\mathrm{c}$ was estimated and shown to be perfectly
consistent with the estimate of a previous study directly in terms of
the original variables \cite{bittnerjanke}.  Unfortunately, when
considering percolation observables, such as whether a vortex loop
percolates the lattice, slight but statistically significant deviations
from $\beta_\mathrm{c}$ were found.  For all observables considered, the
percolation threshold $\beta_\mathrm{p} > \beta_\mathrm{c}$.  That is,
from these observables one would conclude that the vortices proliferate
too early at a temperature below the critical one (as do geometrical
clusters in 3D).  Yet, when taking the percolation threshold as an
adjustable parameter, reasonable estimates were obtained from
percolation observables for the critical exponents $\nu$ and $\beta$,
consistent with those of the XY model.  The problem with the percolation
threshold is quite possibly related to the way vortex loops are traced
out.  When two vortex segments enter a unit cell, it is not clear how to
connect them with the two outgoing segments.  A popular choice is to
randomly connect them, but it might well be that the resulting network
is too extended and consequently percolates too early.  It is in our
mind conceivable that a proper prescription for connecting vortex
segments could lead to a vortex percolation threshold right at the
critical temperature, in the spirit of the FK construction.

As a final remark, we note that even in cases where no thermodynamic
phase transition takes place, the notion of vortex proliferation can be
useful in understanding the phase structure of the system under
consideration.  An example is provided by the 3D Abelian Higgs lattice
model with compact gauge field \cite{compact}.  In addition to vortices,
the compact model also features magnetic monopoles as topological
defects.  It is well established that in the London limit, where the
amplitude of the Higgs field is frozen, it is always possible to move
from the Higgs region into the confined region without encountering
thermodynamic singularities \cite{FradkinShenker}.  Nevertheless, the
susceptibility data for various observables define a precisely located
phase boundary.  Namely, for sufficiently large lattices, the maxima of
the susceptibilities at the phase boundary do not show any finite-size
scaling.  Moreover, the susceptibility data obtained on different
lattice sizes collapse onto single curves without rescaling, indicating
that the infinite-volume limit is reached.  In Ref.~\cite{compact} it
was argued that this phase boundary marks the location where the
vortices proliferate.  A well-defined and precisely located phase
boundary across which geometrical objects proliferate, yet thermodynamic
quantities remain nonsingular has become known as a \textit{Kert\'esz
line}.  Such a line was first introduced in the context of the Ising
model in the presence of an applied magnetic field \cite{kertesz}.

\acknowledgments 
W.J. would like to thank the organizers of this conference for their
warm hospitality.  A.S. is indebted to Professor H. Kleinert for the
kind hospitality at the Freie Universit\"at Berlin.  This work was
partially supported by the Deutsche Forschungsgemeinschaft (DFG) under
grant No.~JA~483/17-3 and the EU RTN-Network `ENRAGE': {\em Random
Geometry and Random Matrices: From Quantum Gravity to Econophysics\/}
under grant No.~MRTN-CT-2004-005616.

\end{document}